\documentclass[12pt,twoside]{article}
\usepackage{fleqn,espcrc1,epsfig,wrapfig}

\usepackage{mathptmx}
\def\al{\alpha}
\newcommand{\nn}{\nonumber}
\newcommand{\be}{\begin{equation}}
\newcommand{\ee}{\end{equation}}
\newcommand{\bea}{\begin{eqnarray}}
\newcommand{\eea}{\end{eqnarray}}

\begin {document}

\title{Learning about the chiral structure of 
the proton from the hyperfine splitting\thanks{Supported in part by
MCyT and Feder (spain), FPA2001-3598, by CIRIT (Catalonia),
2001SGR-00065 and by EURIDICE, HPRN-CT2002-00311.}}

\author{
A. Pineda\address
{Dept. d'Estructura i Constituents de la Mat\`eria,
  U. Barcelona \\ Diagonal 647, E-08028 Barcelona, Catalonia, Spain}
}

\maketitle

\bigskip

\begin{abstract} 

The complete analytical $O(m_{l}^3\al^5/m_p^2\times (\log{m_\pi},
\log{\Delta}, \log{m_l}))$ contribution to the Hyperfine splitting is
given. It can explain about 2/3 of the difference between experiment
and the pure QED prediction when setting the renormalization scale at
the $\rho$ mass. The suppression of the polarizability piece with
respect the Zemach one seems to be, to a large extent, a numerical
accident. We give an estimate of the matching coefficient of the
spin-dependent proton-lepton operator in heavy baryon effective
theory.

\end{abstract}

\bigskip

High precision measurements in atomic physics provide with a unique
place to determinate some hadronic parameters related with the proton
elastic and inelastic electromagnetic form factors, like the proton
radius and magnetic moment, polarization effects, etc.... One
complication in this program comes from the fact that widely separated
scales are involved in these physical processes. Therefore, it becomes
important to relate the physics at these disparate scales in a model
independent way. Effective field theories (EFT's) are a natural
approach to this problem.  In particular, we need an EFT 
at atomic scales. We will use potential NRQED \cite{pNRQED}. Its
effective Lagrangian for QED weakly bound systems at $O(m\alpha^5)$
can be found in Ref. \cite{pos}. It basically reduces to a
Schr\"odinger equation interacting with ultrasoft photons. Here we are
concerned with the hadronic contributions to this EFT
(see \cite{HF}). They are encoded in the matching coefficients,
i.e. in the potentials. Moreover, we will focus on the logarithmically
enhanced hadronic effects to the hyperfine splitting, which, at the
order of interest, only appear in the delta potential\footnote{We will
not consider here the hadronic effects inherited from the anomalous
magnetic moment of the proton, $\mu_p$, due to the tree level
potential:
\be 
\delta V = {4\pi\al(1+\mu_p) \over 3 m_pm_l}{\bf
S}^2\delta^{(3)}({\bf r}) \,,  
\ee 
since $\mu_p$ is known with very high precision from other sources.}:  
\be 
\delta V = 2{c^{pl}_{4,NR} \over
m_p^2}{\bf S}^2\delta^{(3)}({\bf r}) \,.  
\ee 
The point here is to
obtain the coefficient $c^{pl}_{4,NR}$ from QCD in a controlled way.  In
practice, what one can do is to compute its chiral structure due to 
energies of $O(m_\pi)$ and to parameterize the effects due to energies
of $O(m_\rho)$ with matching coefficients inherited from Heavy Baryon
Effective Theory \cite{HBET}.

\begin{figure}[h]
\vspace*{-2ex}
\makebox[1.0cm]{\phantom b}
\put(-30,25){\epsfxsize=6truecm \epsfbox{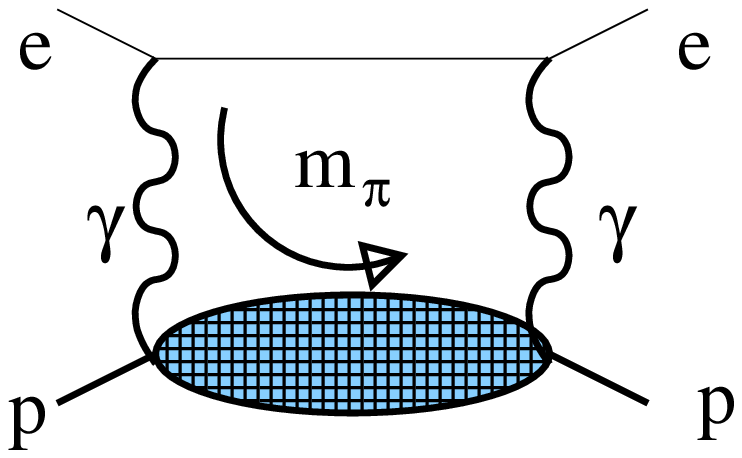}}
\put(210,10){\epsfxsize=6.4truecm \epsfbox{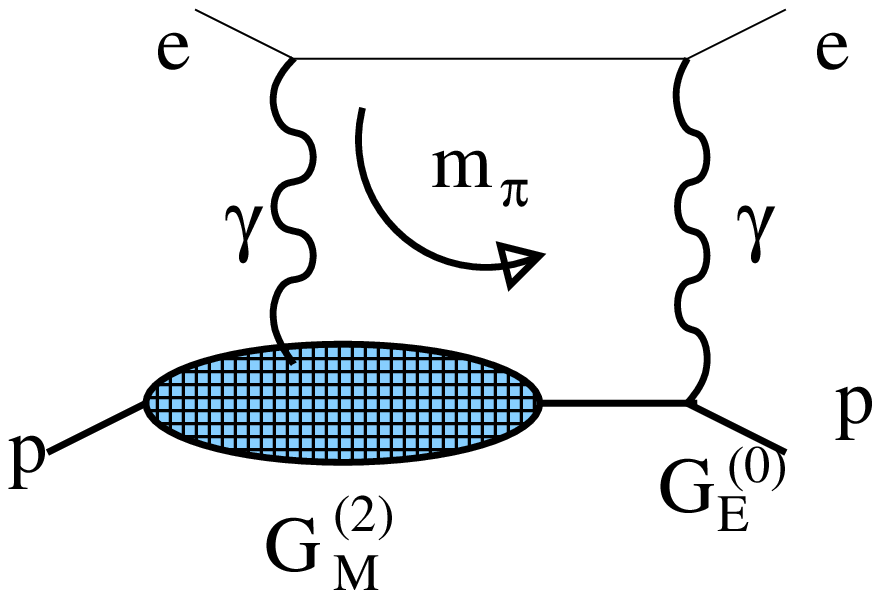}}
\put(-15,15){$a)$}
\put(220,15){$b)$}
\vspace*{-7ex}
\caption {{\it Figure a) corresponds to Eq. (\ref{c4}). Figure b) corresponds to 
Eq. (\ref{Zemachc4}).}}
\label{combined}
\vspace*{-3ex}
\end{figure}

The first non-vanishing contribution to $c_{4,NR}$ appears at 
$O(\al^2)$. Its leading order expression reads (an infrared cutoff larger than 
$m_l\al$ is understood and the
expression for the integrand should be generalized for an eventual full
computation in $D$ dimensions)
\be
\label{c4}
c_{4,NR}^{pl}=-{i g^4 \over 3}\int {d^Dk \over (2\pi)^D}{1 \over k^2}{1 \over
k^4-4m_{l}^2k_0^2 }
\left\{
A_1(k_0,k^2)(k_0^2+2k^2)+3k^2{k_0 \over m_p}A_2(k_0,k^2)
\right\}
\,,
\ee
consistent with the expressions obtained long ago as in Ref. \cite{DS}. It is 
symbolically depicted in Fig. 1a). The bubble is meant to represent the hadronic 
structure of the proton, which, at the order of interest, is represented by the 
forward virtual-photon Compton tensor (needed at $O(1/F_0^2)$
\cite{JO,HF} for the spin-dependent terms),
\begin{equation}  \label{forw-amp2}
 T^{\mu\nu} = i\!\int\! d^4x\, e^{iq\cdot x}
  \langle {p,s}| {T J^\mu(x)J^\nu(0)} |{p,s}\rangle
\,,
\end{equation}
which has the following structure ($\rho=q\cdot p/m$):
\bea \label{inv-dec}
 T^{\mu\nu} =
  &&\left( -g^{\mu\nu} + \frac{q^\mu q^\nu}{q^2}\right) S_1(\rho,q^2) 
   + \frac1{m_p^2} \left( p^\mu - \frac{m_p\rho}{q^2} q^\mu \right)
    \left( p^\nu - \frac{m_p\rho}{q^2} q^\nu \right) S_2(\rho,q^2) \nn \\
  && - \frac i{m_p}\, \epsilon^{\mu\nu\rho\sigma} q_\rho s_\sigma A_1(\rho,q^2)
     - \frac i{m_p^3}\, \epsilon^{\mu\nu\rho\sigma} q_\rho
   \bigl( (m_p\rho) s_\sigma - (q\cdot s) p_\sigma \bigr) A_2(\rho,q^2).
\eea
In many cases only the so called Zemach correction 
\cite{Zemach} is considered.  It corresponds to the Born
approximation of the above expressions (see Fig. 1b)). It reads (for
the definitions see \cite{HF})
\be
\label{Zemachc4}
\delta c_{4,Zemach}^{pl}= 
(4\pi\alpha)^2m_p{2 \over 3}\int {d^{D-1}k \over (2\pi )^{D-1}}
{1 \over {\bf k}^4}G_E^{(0)}G_M^{(2)}
\,.
\ee
It turns out to be the dominant contribution even if, formally, the
other pieces are equally important from the power counting point of
view. 

The total sum in the SU(2) case reads (including point-like, Zemach
and polarizability effects) \cite{HF}\footnote{Note that a dipole
parameterization of the form factors (see, for instance, \cite{BY,KS}) would
be unable to give the chiral logs, since they do not incorporate the
correct non-analytical behavior in the momentum dictated by the chiral
symmetry.}
\bea
\label{Zemachud}
\delta c_{4,NR}^{pl}
&\simeq&
\left(1-{\mu_p^2 \over 4}\right)\alpha^2\ln{m_{l}^2\over \nu^2}
+{b_{1,F}^2 \over 18}\alpha^2\ln{\Delta^2 \over \nu^2}  
 +
{m_p^2 \over (4\pi F_0)^2}\alpha^2{2 \over 3}
\left({2 \over 3}+{7 \over 2\pi^2}\right)\pi^2g_A^2\ln{m_\pi^2 \over \nu^2}
\nn
\\&&
+
{m_p^2 \over (4\pi F_0)^2}\alpha^2{8 \over 27}
\left({5 \over 3}-{7 \over \pi^2}\right)\pi^2g_{\pi N \Delta}^2
\ln{\Delta^2 \over \nu^2}
\,,
\eea
where we have used (the definition of $C$ can be found in the Appendix
of Ref. \cite{HF}) 
\be
C ={\pi^3 \over 12}-{7 \over 8}\pi
=-0.165037
\,,
\ee
for the polarizability corrections (see \cite{HF}). 
We can see that there appears to be a numerical cancellation compared
with the natural size of each term in $C$. This is consistent with the
fact that, experimentally, polarizability corrections seem to be small
\cite{HFfitting}. It is also interesting to consider the large $N_c$ limit of
Eq. (\ref{Zemachud}), it reads (we neglect $\ln(\Delta/m_\pi)$ terms for
consistency) 
\be
\label{Nc}
\delta c_{4,NR}^{pl} (N_c \rightarrow \infty)
\simeq
\alpha^2\ln{m_{l}^2\over \nu^2}
 +
{m_p^2 \over (4\pi F_0)^2}\alpha^2
\pi^2g_A^2\ln{m_\pi^2 \over \nu^2}
\,.
\ee

With Eq. (\ref{Zemachud}), one can obtain the leading hadronic
contribution to the hyperfine splitting. It reads ($\mu_{lp}$ is the
reduced mass)
\be
E_{\rm HF}=4{c^{pl}_{4,NR} \over m_p^2}{1 \over \pi} (\mu_{lp}\al)^3
\,.
\ee
By fixing the scale $\nu=m_\rho$ we obtain the following number for
the total sum in the SU(2) case:
\be
\label{HFtotal}
E_{\rm HF,logarithms}(m_\rho)=- 0.031\;{\rm MHz}
\,.
\ee 
Equation (\ref{HFtotal})
 accounts for approximately 2/3 of the difference between theory (pure QED)
\cite{BY} and experiment \cite{exp}. 
What is left gives the expected size of the counterterm.
Experimentally what we have is $c^{pl}_{4,NR}=-48\alpha^2$ and
$c^{pl}_{4,R}(m_\rho)\simeq c^{p}_{4,R}(m_\rho)\simeq -16\alpha^2$. This last figure gives the
 expected size
of the matching coefficient that appears in the heavy
baryon effective Lagrangian: 
\be
\label{LNl}
\delta {\cal L}_{(N,\Delta)l}=\displaystyle\frac{1}{m_p^2}\sum_l c_{4,R}^{pl}{\bar N}_p \gamma^j \gamma_5
N_p 
  \ \bar{l}\gamma_j \gamma_5 l
\,.
\ee

\end{document}